\documentclass[aps,prl,amsmath,amssymb,onecolumn,nofootinbib,showpacs,tightenlines,12pt]{revtex4}

\usepackage{epsfig,graphicx}
\usepackage{bm}
\newcommand{\be}{\begin{equation}}
\newcommand{\ee}{\end{equation}}
\newcommand{\bea}{\begin{eqnarray}}
\newcommand{\eea}{\end{eqnarray}}

\begin{document}
\title{Geodesic stability for  KS Black hole in Ho$\check{\textbf{r}}$ava-Lifshitz
 gravity via Lyapunov exponents }
 \author{MR.Setare}
 \affiliation{Department of Science, Payame Noor University, Bijar,
Iran} \email{rezakord@mail.ipm.ac.ir}
\author{D.Momeni}
\affiliation{Department of Physics,faculty of Basic Sciences,Tarbiat
Moalem University,Karaj, Iran}

 \pacs{ 3.70.+k, 11.10.-z ,
11.10.Gh, 11.10.Hi }
 \pacs{ 04.70.Bw, 04.50.Gh}
\begin{abstract}

By computing the Lyapunov exponent, which is the inverse of the
instability time scale associated with this geodesic motion we show
that there is two region of space which in both of them the
equatorial  timelike geodesics are stable via Lyapunov measure of
stability.
 \end{abstract} \maketitle
\section{ Mathematical Preliminaries:LYAPUNOV EXPONENTS}
Let us consider a trajectory (in phase space) described by a certain
evolution. The Lyapunov exponents (also known as characteristic
exponents) associated with a trajectory are essentially a measure of
the average rates of expansion and contraction of trajectories
surrounding it. They are asymptotic quantities, defined locally in
state space, and describe the exponential rate at which a
perturbation to a trajectory of a system grows or decays with time
at a certain location in the state space[1]. Analysis conducted with
Lyapunov exponents are called Lyapunov stability analysis. They are
useful in characterizing the asymptotic state of an evolution
(attractors in dissipative systems)[2]. Using Lyapunov exponents, we
can distinguish among fixed points, periodic motions, quasiperiodic
motions, and chaotic motions[3].
\subsection{Concept of Lyapunov Exponents}
We begin by defining Lyapunov exponents for a given system of
equations. Let $X(t)$ such that $X(t = 0) = X_{0}$ represent a
trajectory of the system governed by the following n-dimensional
autonomous system:
\begin{eqnarray}
\dot{x}=F(x;M)
\end{eqnarray}
where the vector $x$ is made up of $n$ state variables, the vector
function F describes the nonlinear evolution of the system, and M
represents a vector of control parameters. Denoting the perturbation
provided to $X(t)$ by $y ( t )$ and assuming it to be small, we
obtain an equation after linearization in the disturbance terms. The
perturbation is governed by
\begin{eqnarray}
\dot{y}=Ay
\end{eqnarray}
where, in general, $A = D_{x}F[x(t);M]$ is an $n\times n$ matrix
with time dependent coefficients. If we consider an initial
deviation $y(O)$, its evolution is described by
\begin{eqnarray}
y(t)=\phi(t)y(0)
 \end{eqnarray}
  where $\phi(t)$ is the fundamental (transition) matrix solution of (3)
associated with the trajectory $X(t)$.The eigenvalues of A provide
information about the stability of the associated fixed point.\\
The procedure used to determine Lyapunov exponents can be considered
to be a generalization of linear stability analyses. An interesting
and detailed discussion on the relationship between linear stability
analysis and Lyapunov stability analysis can be found in the paper
of Goldhirsch, Sulem, and Orszag [4]. They argue that the Lyapunov
exponents are global quantities associated For an appropriately
chosen $y(0)$ in (3), the rate of exponential expansion or
contraction in the direction of $y(0)$ on the trajectory passing
through $X_{0}$ is given by
\begin{eqnarray}
\lambda_{j}=lim_{t\rightarrow\infty}\frac{1}{t}\log{\frac{||y(t)||}{||y(0)||}}
\end{eqnarray}

where the symbol $||$ denotes a vector norm . The asymptotic
quantity $\lambda_{j}$ is called the Lyapunov exponent. We have n
Lyapunov exponents associated with an n-dimensional autonomous
system. It can be shown that if the trajectory $X ( t )$ corresponds
to a motion other than a fixed point, then one of the $\lambda_{j}$
is always zero . Following Lyapunov [5], the fundamental matrix
$\phi(t)$ is called regular if
\begin{eqnarray}
lim_{t\rightarrow\infty}\frac{1}{t}\log{det(\phi(t))}
\end{eqnarray}
exists and is finite. If $\phi(t)$ is regular, then, according to a
theorem , the asymptotic quantity defined in (4) exists and is
finite for any initial deviation y(0) belonging to the n-dimensional
space.
\section{stability of circular orbits via Lyapunov exponents }
As was shown by Cardoso et al[6 ] for all spherically symmetric
spacetimes, in a geometrical optics approximation, QNMs can be
interpreted as particles trapped at unstable circular null geodesics
and slowly leaking out. The leaking time scale is given by the
principal Lyapunov exponent, which one can  obtain a fairly simple
expression, in terms of the second derivative of the effective
radial potential for geodesic motion. As was stated by Cardoso et
al, this deep, intuitive approach to the QNMs  and it's relation to
the circular null geodesics is valid for all asymptotically flat,
spherically symmetric black-hole spacetimes.The most important note
is that the powerfull formalism which is presented in [5] is
completely independent from the form of the action, i.e there is no
difference between treating a stationary, asymptotically flat,
static spacetime from a higher order gravity as f(R) or HL and the
usual higher dimensional (or 4-dim) BHs in GR.Thus with have no
worrying about the validity of their method we can use it and
following all the steps of the [6].Without no loss of generality we
can restrict ourselves to a simple kind of problems includes
circular orbits in stationary spherically symmetric spacetimes and
equatorial circular orbits in stationary spacetim .This case could
be described by a two-dimensional phase space which for our
geodesics analysis may be described by
\begin{eqnarray}
X(t)=(p_{r},r)
\end{eqnarray}
Linearizing the equations of motion  about orbits of constant r ,
the principal Lyapunov exponents can be expressed as
\begin{eqnarray}
\lambda=\sqrt{\frac{V_{r}''}{2\dot{t}^{2}}}
\end{eqnarray}
Where $V_{r}=\dot{r}^{2}$ and  a dot denotes the derivative with
respect to  the proper time $\tau$. Remember to mind that For
circular geodesics
\begin{eqnarray}\nonumber
V_{r}=V'_{r}=0
\end{eqnarray}

\section{KS black hole solution in HL theory}
There are different versions of the HL theory.As a geometrical point
of view all these versions  Follow from the ADM decomposition of the
metric [7],  the fundamental objects of interest are the fields
$N(t,x),N_{i}(t,x),g_{ij}(t,x)$ corresponding to the \emph{lapse },
\emph{shift} and \emph{spatial metric} of the ADM decomposition.
 In the $(3 + 1)$-dimensional ADM formalism, where the
metric can be written as
 \begin{eqnarray}\nonumber
ds^2=-N^2dt^2+g_{ij}(dx^{i}+N^{i}dt)(dx^{j}+N^{j}dt)
\end{eqnarray}
and for a spacelike hypersurface with a fixed time, its extrinsic
curvature $K_{ij}$ is

\begin{eqnarray}\nonumber
K_{ij}=\frac{1}{2N}(\dot{g_{ij}}-\nabla_{i}N_{j}-\nabla_{j}N_{i})
\end{eqnarray}
where a dot denotes a derivative with respect to $t$ and covariant
derivatives defined with respect to the spatial metric $g_{ij}$, the
action of Ho$\check{\textbf{r}}$ava-Lifshitz theory  for $z=3$ is
\begin{eqnarray}\nonumber
S=\int_{M} dtd^{3}x\sqrt{g} N(\mathcal{L}_{K} - \mathcal{L}_{V} )
\end{eqnarray}
we define the space-covariant derivative on a covector $v_{i}$ as
$\nabla_{i}v_{j}\equiv \partial_{i}v_{j}-\Gamma_{ij}^{l}v_{l}$ where
$\Gamma_{ij}^{l}$ is the spatial Christoffel symbol. $g$ is the
determinant of the 3-metric and $N = N(t)$ is a dimensionless
homogeneous gauge field. The kinetic term is
\begin{eqnarray}\nonumber
\mathcal{L}_{K}=\frac{2}{\kappa^2}\mathcal{O}_{K}=\frac{2}{\kappa^2}(K_{ij}K^{ij}-\lambda
K^2)
\end{eqnarray}
Here $N_{i} $ is a gauge field with scaling dimension $[N_{i}] = 2$.\\
The \emph{'potential'} term $\mathcal{L}_{V}$ of the
$(3+1)$-dimensional theory is determined by the \emph{principle of
detailed balance }[8], requiring $\mathcal{L}_{V}$ to follow, in a
precise way, from the gradient flow generated by a 3-dimensional
action $W_{g}$. This principle was applied to gravity with the
result that the number of possible terms in $\mathcal{L}_{V} $ are
drastically reduced with respect to the broad choice available in an
'\emph{potential} is
\begin{eqnarray}\nonumber
\mathcal{L}_{V}=\alpha_{6}C_{ij}C^{ij} - \alpha_{5}\epsilon_{l}^{ij}
R_{im}\nabla_{j}R^{ml} + \alpha_{4} [R_{ij}R^{ij}-
\frac{4\lambda-1}{4(3\lambda-1)} R^2] +\alpha_{2}(R - 3\Lambda_{W})
\end{eqnarray}
Where in it $C_{ij}$ is the \emph{Cotton }tensor [9 ]which is
defined as,
\begin{eqnarray}\nonumber
C^{ij}=\epsilon^{kl(i}\nabla_{k}R^{j)}_{l}
\end{eqnarray}
The kinetic term could be rewritten in terms of the \emph{de Witt
metric} as:
\begin{eqnarray}\nonumber
\mathcal{L}_{K}=\frac{2}{\kappa^2}K_{ij}G^{ijkl}K_{kl}
\end{eqnarray}
Where we have introduced the \emph{de Witt metric}
\begin{eqnarray}\nonumber
G^{ijkl}=\frac{1}{2}(g^{ik}g^{jl}+g^{il}g^{jk})-\lambda g^{ij}g^{kl}
\end{eqnarray}

Inspired by methods used in quantum critical systems and non
equilibrium critical phenomena, Ho$\check{\textbf{r}}$ava restricts
the large class of possible potentials using the principle of
detailed balance outlined above. This requires that the potential
term takes the form
\begin{eqnarray}\nonumber
\mathcal{L}_{V}=\frac{\kappa^2}{8}E^{ij}G_{ijkl}E^{kl}
\end{eqnarray}
Note that by constructing $E^{ij} $ as a functional derivative it
automatically transverse within the foliation slice,
$\nabla_{i}E^{ij}=0$. The equations of motion were  obtained in
[10]. KS BH is a static spherically symmetric solution for HL theory
which contains 2 parameter , one mass like parameter  $m$ and a
parameter which controls the escape from a naked singularity
$\omega$ and satisfies [11]
\begin{eqnarray}\nonumber
\omega m^{2}\geq\frac{1}{2}
\end{eqnarray}
In the usual spherical coordinates  $(t,r,\theta,\phi)$ and in the
Schwarzschild's gauge the metric reads:
\begin{eqnarray}
ds^{2}=diag(+f,-\frac{1}{f},-r^{2}\Sigma_{2})
\end{eqnarray}
where in it the metric gauge function is
\begin{eqnarray}
f=1+\omega r^{2}-\sqrt{\omega^2 r^{4}+4 m \omega r}
\end{eqnarray}
and is  $\Sigma_{2}$ the surface element on a unit 2- sphere. As
motivated by Sekiwa "\emph{it is obvious that $1/2\omega$ is
equivalent to $Q^{2}$ and this means that we could view $1/2\omega$
as a charge in some degree}"[12]. Thus The outer and inner event
horizon can be compared with the outer and inner event horizon of
Reissner-Nordstrom black hole [13]. Essentially as claimed by the
founders of the KS, this solution "\emph{represents the analog of
the Schwarzschild solution of GR}".
\section{Circular orbits}
The best treatment of the geodesics equations is due to
Chandrasekhar [14].For a simple typical form of a spherically
symmetric metric in Schwarzschild gauge
\begin{eqnarray}
ds^{2}=diag(+f(r),-\frac{1}{g(r)},-r^{2}\Sigma_{2})
\end{eqnarray}
one can obtain the following expression for the potential function:
\begin{eqnarray}
V_{r}=g(r)(\frac{E^{2}}{f(r)}-\frac{L^{2}}{r^{2}}-\delta_{1})
\end{eqnarray}
Where in it the $E,L$ respectably can be interpreted as the
Energy,angular momentum of the test particle within the circular
orbits in similar to the classical mechanics and $\delta_{1}=1,0$
for timelike and null geodesics, respectively.For KS BH (8) the
relation (11) convert to:
\begin{eqnarray}
V_{r}=E^{2}-\frac{L^{2}f(r)}{r^{2}}-\delta_{1}f(r)
\end{eqnarray}
\section{Timelike geodesics}
For circular orbits we know that both the potential term and the
first derivates of it must be vanish which leads us to the next
expression for the second derivative of the potential
\begin{eqnarray}
V''_{r}=2[\frac{-3ff'/r+2f'^{2}-ff''}{2f-rf'}]
\end{eqnarray}
thus using (7) the Lyapunov exponent at the circular timelike
geodesics is
\begin{eqnarray}
\lambda=\frac{1}{2}\sqrt{(2f-rf')V''_{r}}
\end{eqnarray}
Since the energy must be real, we require
\begin{eqnarray}
\frac{\partial\log(f(r))}{\partial\log(r)}<2
\end{eqnarray}
this inequality leads to the next bound for the radius of the
timelike circular orbit\footnote{Remember that in KS BH we must have
$m^{2}\omega\geq\frac{1}{2}$.}
\begin{eqnarray}
\omega^{2}r^{3}-9m^{2}\omega^{2}r+4m\omega>0
\end{eqnarray}
This is a cubic algebraic equation . We review the general solution
for a cubic eq.Let the general cubic eq be
\begin{eqnarray}
x^{3}+px+q=0
\end{eqnarray}
To have three distinct real root(the case which is happen in KS BH)
we define
\begin{eqnarray}
\Delta=4p^{3}+27q^{2}
\end{eqnarray}
If $\Delta<0$ then these three roots can be obtained by solving a
simple trigeometric
 \begin{eqnarray}
 \cos(3\theta)=\cos(\alpha)
 \end{eqnarray}
 where in it
 \begin{eqnarray}
 \cos(\alpha)=-\frac{4q}{A^{3}}\\\nonumber
 A^{2}=-\frac{4p}{3}\\\nonumber
 x=A\cos(\theta)
 \end{eqnarray}
 for radial eq (16) we have
 \begin{eqnarray}
 \Delta=-108(\frac{m}{\omega})^{2}\delta\leq -297(\frac{m}{\omega})^{2}
 \end{eqnarray}
where in it
\begin{eqnarray}
\delta=27 m^{6}\omega^{6}-4m^{2}\omega^{4}\geq 2.75
\end{eqnarray}
finally we can write the next expressions for all roots of  the (16)
\begin{eqnarray}
r_{n}=2\sqrt{3}m\cos(\frac{2n\pi}{3}\pm\frac{\alpha}{3}),n=0,1,2
\end{eqnarray}
Also we know that the KS solution has a physically accessible
horizon located at
\begin{eqnarray}
r=h=m+\sqrt{m^{2}-Q^{2}}
\end{eqnarray}
The KS solution is valid only for such region of the space where
$r\geq h$. We must determine that which of these three distinct real
values $r_{i},i=0,1,2$ belongs to this interval.Indeed we must have
\begin{eqnarray}
h<r_{n}\leq2\sqrt{3}m
\end{eqnarray}
(Which is satisfied automatically)

Also the negativity of the energy impose that for timelike geodesics
we must restricted ourselves to
\begin{eqnarray}
(r_{0}<r<r_{1})\curlyvee(r>r_{2})
\end{eqnarray}
This bound for the radial coordinates which arisen from the reality
of the Energy has an essential role for investigating the stability
via the relation (14) for the Lyapunov exponent.If we want to have a
stable motion the exponent $\lambda$ must be imaginary and have a
negative real part and unstablee unless.For instability we must have

\begin{eqnarray}
V''_{r}>0
\end{eqnarray}
The dominator of it must be checked.That is we must check that when
r belongs to the (26) the next inequality satisfied or not?
\begin{eqnarray}
-3ff'/r+2f'^{2}-ff''>0
\end{eqnarray}
the opposite sign indicates on astable one. A very simple and
straightforward calculation show that for all values of the r belong
to the (26) we have
\begin{eqnarray}
V''_{r}<0
\end{eqnarray}
Thus the time like geodesics in KS BH are stable.
\section{conclusion}
HL theory brings some important new features from the GR to the
higher dimensional lagrangian and it's role in construction a non
relativistic candidate for quantum gravity.The Lyapunov exponents is
a very important and powerful method to stability any dynamical
system both in classical mechanics and also in quantum treatment of
complex systems.I n this short letter after a brief review of the
Lyapunov method in mathematical physics, following the method which
have been described by Cardoso et al[5] for 4 and 5 dimensional BHs
in GR,we applied a similar method to the a new type of the BH which
indeed are a formal IR limit of the HL theory and show that the
timelike geodesics are stable under small perturbations.As Cardoso
et al show that for a equatorial circular timelike geodesics in a
Myers-Perry black-hole background are unstable .But we show that in
the context of the  HL  theory these geodesics are  stable.Thus we
can argued that the nonlinear terms are successful to pass from the
instabilities.

\end{document}